\documentclass[aps,pra,reprint,groupedaddress, showpacs,amssymb,nofootinbib,]{revtex4-1}
\usepackage{amsmath,amssymb}
\usepackage{bm}
\usepackage{epsfig}
\numberwithin{equation}{section}

\usepackage{enumerate}
\usepackage{braket}
\usepackage{relsize}
\usepackage{hyperref}
\hypersetup{
colorlinks = true,
linkcolor = blue,
citecolor = blue,
urlcolor = blue, 
}
\everymath{\displaystyle}

\def\cf{{\it cf.~}}
\begin{document}

\title{Decoherence in the classical limit of histories of a particle coupled to a von Neumann apparatus}

\author{Francesc S. Roig}
\email{roig@physics.ucsb.edu}

\affiliation{Physics Department University of California Santa Barbara, CA 93106, USA}

\date{\today}


\begin{abstract}
{\label{abstract}}
Using the Gell-Mann and Hartle formalism of generalized quantum mechanics of closed systems, we study the classical limit of coarse-grained spacetime histories and their decoherence. The system under consideration is one-dimensional and consists of a particle coupled to a von Neumann apparatus that performs a measurement of the position of the particle during the finite time interval during which the histories of this system take place.  We consider two cases: a free particle and a harmonic oscillator. The real line is divided into intervals of the same length, and coarse-grained histories are defined by the time average of the position of the particle on a given Feynman path to be within one of these intervals. The position of the pointer in each Feynman path correlates with this time average. The class operators for this system have been evaluated, and the decoherence functional shows that these coarse-grained histories do not decohere, not even when initially either the particle or the pointer is in an eigenstate of position.  Decoherence is obtained only when the classical limit is taken.  Qualitative arguments for decoherence in the classical limit are presented for the case of a general particle potential. 
\end{abstract}

\pacs{03.65.Ta}

\maketitle


\section{Introduction}\label{Introduction}


In the current work we set out to investigate decoherence in the sense of Refs.~\cite{H1,H2}, when applied to a system consisting of a particle whose position is measured by a von Neumann apparatus during a finite time interval.  For each Feynman path, the indication of the apparatus will correlate with the weighted time average of the position of the particle.  That is
\begin{equation}\label{weighted-bar x}                         
\bar x[x(t),T]=\frac{1}{T}\int_0^T f(t) x(t){\rm d}t,
\end{equation}
 as shown in Ref.~\cite{FR-2}.  The weight function~$f(t)$ is a dimensionless coupling function of time  between the particle and the apparatus with compact support~$[0,T]$.
 
We reproduce the steps from Ref.~\cite{FR-2} where
the Hamiltonian for this system is
\begin{equation} \label{H}
\hat H=\frac{\hat p^2}{2m}+V(x,t)+\frac{\hat P^2}{2M}+\hat H_i.
\end{equation}
The mass of the particle is~$m$ and the mass of the apparatus or pointer is~$M$. The total potential acting on the particle is~$V(x,t)$.

The interaction between the particle and the apparatus is described by the von Neumann Hamiltonian introduced in Ref.~\cite{vonN},
\begin{equation} \label{Hi}
\hat H_i = \frac{1}{T} f(t) x\hat P,
\end{equation} 
where
\begin{equation}\label{pointer_momentum}
\hat P=-{\rm i}\hbar\frac{\partial}{\partial X}
\end{equation}
 is the momentum of the pointer,with~$X$ the pointer position, or indicator variable. The duration of this interaction is~$T$.  Just before the measurement, the system is described by the pure state $\psi_0(x, X) = \varphi_0(x) \Phi_0(X)$, where $\varphi_0(x)$ is the wavefunction for the state of the particle and~$\Phi_0(X)$ can be a narrow wavepacket describing the apparatus.
 
 The criterion for coarse-grained histories will be established as follows: The real line will be divided up into contiguous intervals of the same length, and a coarse-grained history will consist of the  set of all Feynman histories in the time interval~$[0,T]$ whose weighted time average as defined by Eq.~\eqref{weighted-bar x} lies in one of those intervals.   Unlike  Ref.~\cite{FR-2}, we will find that there is no decoherence even in the cases of either sharp initial states of position for the particle or for the pointer. Decoherence will be obtained only in the classical limit.  Finally, for review works on decoherence in quantum mechanics see references~\cite{{Dowker-Kent},{Schlosshauser},{Dowker-Halliwell},{Halliwell}}.  
 
In Ref.~\cite{FR} it was shown that the propagator for a particle-pointer system  is written as a sum over all pointer and particle paths between~$0$ and~$T$ as
\begin{equation} \label{2D-pathint}
\left\langle {x,X} \right| e^{ - \frac{{\rm i}}{\hbar}\hat HT} \left| {x',X'} \right\rangle  =\int_U\delta X(t) \int_u\delta x(t)e^{\frac{{\rm i}}{\hbar}S[x(t),X(t)]},
\end{equation}
with
\begin{align}\label{allhistories}
U&=\Set{ X(t) \,\big | X(0)=X', \,X(T)=X}\nonumber\\
u&=\Set{ x(t)\, \big | x(0)=x', \,x(T)=x}
\end{align}

and where~$S$ is the action 
\begin{align}\label{S1}
S\left[ {x(t),X(t)} \right] = &\mathlarger{{\mathlarger{\int}}}_{\!\!\!\!\!0}^T\Bigg[ \frac{m}{2}\dot x^2  - V(x,t)\nonumber\\
&  
 +\frac{M}{2}\left( {\dot X - \frac{f(t)}{T}x} \right)^2\Bigg] {\rm d}t.
\end{align}

The path integral over all pointer paths~$X(t)$ in Eq.~\eqref{2D-pathint} can be readily evaluated. To that end we
change the pointer variable according to
\begin{equation}
Y=X-\frac{1}{T}\int_0^t f(t')x(t'){\rm d}t'.
\end{equation}
With this change the double path integral~in \eqref{2D-pathint} can be rewritten
\begin{align} \label{1D-pathint}
&\mathlarger{\iint} \delta x(t)\delta X(t)e^{\frac{\rm i}{\hbar}S[x(t),X(t)]}=\nonumber\\
&\mathlarger{\mathlarger{\int}} \delta x(t)~{\rm exp}\left\{\frac{\rm i}{\hbar}\int_0^T{\rm d}t \left[\frac{m}{2}\dot x^2
-V(x,t)\right] \right\}\nonumber\\
&\times
\mathlarger{\mathlarger{\int}} \delta Y(t)~{\rm exp}\left(\frac{\rm i}{\hbar}\int_0^T {\rm d}t\frac{M}{2}\dot Y^2\right).
\end{align}
The path integration over~$Y(t)$ yields the result for a free particle of mass~$M$
\begin{align}
\mathlarger{\mathlarger{\int}}  \delta &Y(t)~{\rm exp}\left(\frac{\rm i}{\hbar}\int_0^T {\rm d}t\frac{M}{2}\dot Y^2\right)=\nonumber\\
&
\sqrt{\frac{M}{2\pi i\hbar T}}~{\rm exp}\left[\frac{\rm i}{\hbar}\frac{M}{2T}\left(Y'-Y\right)^2\right].
\end{align}
That is, the propagator~\eqref{2D-pathint} can be expressed in terms of a single path integral
over all particle paths
\begin{align}\label{X-pathint}
&\Braket{x,X|{\rm e}^{-\frac{\rm i}{\hbar}HT}|x',X'} =\nonumber\\
&
\mathlarger{\mathlarger{\int}} \delta x(t)~{\rm exp}\left[\frac{\rm i}{\hbar}\int_0^T {\rm d}t\left(\frac{m}{2}\dot x^2-V(x,t)\right)\right]\nonumber\\
&\times 
\sqrt{\frac{M}{2\pi {\rm i}\hbar T}}
~{\rm exp}\left\{\frac{\rm i}{\hbar} \frac{M}{2T}\Big[X-X'-\bar x[x(t),T]\Big]\right\},
\end{align}
where~$\bar x[x(t),T]$ is given by~\eqref{weighted-bar x} .

If initially the particle is in the pure state wavefunction~$\psi_0(x,X)=\varphi_0(x)\Phi_0(X)$, then
the wavefunction for the system at time~$T$ is given by the entangled state
\begin{align}\label{FinalStateT}
\psi&(x,X,T)=\int_{-\infty}^\infty{\rm d}x'\varphi_0(x')\nonumber\\
&\times
\mathlarger{\mathlarger{\int}}_u\delta x(t)
{\rm exp}\left\{\frac{\rm i}{\hbar}\int_0^T\left[\frac{m}{2}\dot x^2-V(x,t)\right]{\rm d}t\right\}\nonumber\\
&\times
\Phi\left(X-\bar x[x(t),T]\,\right),
\end{align}
where
\begin{align}\label{Phi0}
&\Phi\left(X-{\bar x}[x(t),T]\,\right)=\sqrt{\frac{M}{2\pi {\rm i}\hbar}}\nonumber\\
&\times
\mathlarger{\mathlarger{\int}}_{-\infty}^\infty {\rm d}X'\Phi_0(X')
{\rm exp}\left\{\frac{\rm i}{\hbar} \frac{M}{2T}\Big(X-X'-\bar x[x(t),T]\Big)^2\right\}.
\end{align}

Expression~\eqref{Phi0} shows that on a given Feynman path of the particle, the wavepacket describing the pointer at~$t=0$ and centered at~$X=0$ has spread at ~$t=T$ and shifted its center to~$\bar x [x(t),T]$. Thus  the pointer coordinate correlates with the weighted average of the position of the particle on an individual Feynman path.  If the coupling function in the definition~\eqref{Hi} for the interaction between the apparatus and the particle is a constant, then the pointer coordinate correlates with the time average of the position of the particle over a Feynman path.
 
For a particle in an eigenstate of position at~$x_0$ at~$t=0$ the wavefunction of the system at the end of the measurement is
\begin{align}\label{SharpParticleState}
\psi(x,X,T)=&
\mathlarger{\mathlarger{\int}}_{(x_0,0)}^{(x,T)}\delta x(t)
{\rm exp}\Bigg\{\frac{\rm i}{\hbar}\int_0^T\Big[\frac{m}{2}\dot x^2\nonumber\\
&
-V(x,t)\Big]{\rm d}t\Bigg\}
\Phi\left(X-\bar x[x(t),T]\,\right).
\end{align}
 The path integral sums over all particle paths that start at~$x_0$ at~$t=0$ and end at any point on the real line at~$t=T$. 
 
For quadratic potentials, in addition to the correlation of the pointer with the weighted time average over a Feynman path, there is also a correlation with the arithmetic average of the initial and final position of the particle. Thus the pointer will indicate the same reading for all the Feynman paths with a given arithmetic average of the initial and final position of the particle. See Ref.~\cite{FR-2}.  The weighted time average over a Feynman path provides a finer coarse-graining than for the case of graininess by arithmetic average.

In Sections III and IV we will construct the class operators and decoherence functionals introduced in Refs.~\cite{H1,H2} for the case of coarse-grainings characterized by averages on individual Feynman paths for a free particle and for a harmonic oscillator respectively. For the sake of tractability the coupling function in the interaction~\eqref{Hi} between the particle and the apparatus will be taken to be a constant.  We will find that decoherence occurs only when  the classical limit is taken.

Finally, in section V we will examine qualitatively decoherence in the classical limit for the case of a general potential acting on the particle.


\section{General Potential and General Coupling Function}
 As mentioned earlier, in reference~\cite{FR} it was shown that the propagator for a particle-pointer system is written as a sum over all paths for the particle and the pointer between~$0$ and~$T$ \cf Eq.~\eqref{2D-pathint},  
where the action is defined in Eq.~\eqref{S1}.

As done in Ref.~\cite{FR-1} the sum over pointer paths can easily be carried out by inserting two complete sets of eigenstates of the momentum~$P$ of the pointer and normalized according to~$
\left\langle P \right|\left. {P'} \right\rangle  = \delta (P - P')$.
 The propagator~\eqref{2D-pathint} can be rewritten in terms of a reduced propagator as
\begin{eqnarray} \label{propH_P}
\left\langle {x,X} \right|e^{ - \frac{\rm i}{\hbar}\hat HT} \left| {x',X'} \right\rangle  = &&\int_{ - \infty }^\infty  {\frac{dP}{2\pi\hbar }} e^{ \frac{\rm i}{\hbar}[P(X - X') - (P^2/2M)T]}\nonumber\\
&&\times
 \left\langle x \right|e^{ - \frac{\rm i}{\hbar}\hat H_P T} \left| {x'} \right\rangle,
\end{eqnarray}
with reduced Hamiltonian
\begin{equation} \label{H_P}
\hat H_P=\frac{\hat p^2}{2m}+V(x,t)+\frac{f(t)}{T}x P.
\end{equation}
The reduced propagator~$\left\langle x \right|e^{ - i\hat H_P T} \left| {x'} \right\rangle$ for the particle can be expressed as a sum over all particle paths, and the propagator in Eq.~\eqref{propH_P} becomes
\begin{eqnarray} \label{reduced_path_int}
\left\langle {x,X} \right|e^{ - \frac{\rm i}{\hbar}\hat HT} \left| {x',X'} \right\rangle  = &&\int_{ - \infty }^\infty  \frac{dP}{2\pi\hbar } e^{ \frac{\rm i}{\hbar}[P(X - X') -(P^2/2M)T]}\nonumber\\
&&\times
 \int_u\delta x(t) e^{\frac{\rm i}{\hbar}S_P[x(t)]},
\end{eqnarray}
where the reduced action is
 \begin{equation} \label{S_P}
S_P[x(t)]=\int_0^T{L_P}dt
\end{equation}
and~$L_P$ is the reduced Lagrangian
\begin{equation} \label{L_P}
L_P=\frac{m}{2}\dot x^2-V(x,t)-\frac{f(t)}{T}xP.
\end{equation}

Next we will apply the formalism developed in Refs.~\cite{H1,H2} for closed systems.
The basic formalism on how to apply the quantum mechanics of closed systems to a particle coupled to a von Neumann apparatus was presented in Ref.~\cite{FR-2}. The set of all particle paths~$C$ starting on any point on the real line ($x$-axis) at~$t=0$ and ending on any point on the real line ($x'$-axis) at~$t=T$ was grouped into an exhaustive set of mutually exclusive classes.  The real line of all values for the average position~$\bar x$ on a given Feynman path for the particle was divided into equal length intervals:
\begin{align}\label{alternatives}
\Delta_{\alpha}&=\big(\,\bar x_{\alpha}-\delta/2\,,\,\bar x_{\alpha}+\delta/2\,\big],~~{\rm with}~\delta>0\nonumber\\
\mathbb{R}&=\bigcup_{\alpha\in \mathbb{Z}}\Delta_{\alpha}\,,~{\rm with}~~ \bar x_{\alpha+1}-\bar x_\alpha=\delta.
\end{align}
A class of particle paths~$c_{\alpha}$ was defined by the set of all paths~$x(t)$ on the time interval~$[0,T]$  for which on any given path we get the average~$\bar x$ such that~$\bar x\in \Delta_{\alpha}$.
Therefore  the set~$C$ of all paths in the time interval~$[0,T]$  has been divided into an exhaustive set of mutually exclusive classes~$c_{\alpha}$:
\begin{align}
C&=\bigcup_{\alpha\in \mathbb{Z}}c_{\alpha}\nonumber\\
\O&=c_{\alpha}\bigcap c_{\alpha'},~~{\rm with}~\alpha\not =\alpha'.
\end{align}

The average of~$x$ is given by~\eqref{weighted-bar x}.
 Given a potential~$V(x,t)$ acting on the particle, then for each particle path in a class~$c_{\alpha}$, the pointer correlates with~$\bar x[x(t)]\in\Delta_\alpha$ at the time~$T$. Thus the indication of the pointer is somewhere within the interval~$\Delta_\alpha$ at the end of the finite time measurement. 

 Fine-grained histories for this system are given by any particle and pointer paths~$\big(x(t),X(t)\big)$ defined on the time interval~$[0,T]$. The coarse-grained-classes  consist of all the fine-grained histories belonging to the set
\begin{equation}\label{histories}
c_{\alpha}=\Set{\big(x(t), X(t)\big) | \bar x\equiv\frac{1}{T}\int_0^Tf(t)x(t){\rm d}t\in \Delta_{\alpha}},
\end{equation}
with no restriction on the paths~$X(t)$ for the pointer.
 For a given class, each of the fine-grained histories of the particle in this class have the value of their average position to be such that~$\bar x\in \Delta_\alpha$. 
 
The essential feature of the quantum mechanics of closed systems applied to our system is that for each class of paths~$c_{\alpha}$, a class operator~$\widehat C_{\alpha}$ can be defined whose matrix elements in the position representation take the form of a path integral unrestricted over all pointer paths and restricted to all possible particle paths in~$c_{\alpha}$
\begin{equation}\label{class-operator}
\left\langle {x,X} \right|\widehat C_{\alpha}\left| {x',X'} \right\rangle=
\int_U\delta X(t)
\int_{c_{\alpha}}\delta x(t) \,{\rm e}^{\frac{\rm i}{\hbar} S[X(t),x(t)]},
\end{equation}
where the action~$S$ is given by Eq.~\eqref{S1}. 

When the system starts at~$t=0$ in the state described by the state vector~$\Ket{\Psi_0}$, the decoherence functional is defined by
\begin{equation}\label{decoherence-functional}
D(c_{\alpha},c_{\alpha'})=\Braket{\Psi_0|\widehat C_{\alpha'}^{\,\dagger}\,\,\widehat C_\alpha|\Psi_0}.
\end{equation}
The quantum mechanics of closed systems does not predict probabilities for every possible set of coarse-grained histories but only for those that satisfy the decoherence condition
\begin{equation}\label{decoherence}
D(c_{\alpha},c_{\alpha'})\thickapprox 0,\quad\alpha'\not =\alpha.
\end{equation}
In such a case, probabilities for each of the coarse grained classes~$c_{\alpha}$ can be assigned according to  
\begin{equation}
p_\alpha=\parallel\widehat C_{\alpha}\Ket{\Psi_0}\parallel^2.
\end{equation}

From the expression~\eqref{class-operator} we can obtain the analogue of Eq.~\eqref{reduced_path_int} by simply restricting the sum over all paths to only those paths in the class~$c_{\alpha}$.  That is,
\begin{align} \label{class -operator1}  
\left\langle {x,X} \right|\widehat C_{\alpha}\left| {x',X'} \right\rangle =&
 \int_{ - \infty }^\infty  \frac{dP}{2\pi\hbar } e^{ \frac{\rm i}{\hbar}[P(X - X') -(P^2/2M)T]}\nonumber\\
&\times
 \int_{c_{\alpha}}\delta x(t) e^{\frac{\rm i}{\hbar}S_P[x(t)]}.
\end{align}

 The matrix elements of~$\widehat C_\alpha$ in Eq.~\eqref{class-operator} can be rewritten introducing the function
\begin{equation}\label{top-hat function}
\mathlarger{e}_{\mathsmaller{\Delta}_{\alpha}}(\bar x)=
\begin{cases}
\,\,1,\quad \bar x\in\Delta_\alpha\\
\,\,0,\quad \bar x\notin\Delta_\alpha\,,
\end{cases}
\end{equation}
where~$\bar x$ is defined by \eqref{weighted-bar x}.
Thus we may rewrite the functional integral over the class of paths~$c_\alpha$ in Eq.\eqref{class-operator} as
\begin{align}\label{particle-path-int1}
\int_{c_{\alpha}}&\delta x(t)\,{\rm e}^{\frac{\rm i}{\hbar} S_P[x(t)]}=\nonumber\\
&
\mathlarger{\int}_u\delta x(t)\mathlarger{e}_{\mathsmaller{\Delta}_{\alpha}}\left(\bar x\right){\rm e}^{\frac{\rm i}{\hbar}S_P[x(t)]}.
\end{align}
  When we introduce the Fourier transform
\begin{equation}\label{fourier transform}
\mathlarger{e}_{\mathsmaller{\Delta}_{\alpha}}(\bar x)=\frac{1}{\sqrt{2\pi}}\int_{-\infty}^\infty {\rm
 d}k e^{{\rm i}k\bar x}\mathlarger{\tilde e}_{\mathsmaller{\Delta}_{\alpha}}(k)
 \end{equation}
 and following Ref.~\cite{B-H}, the expression for~$\mathlarger{\tilde e}_{\mathsmaller{\Delta}_{\alpha}}(k)$ can be evaluated explicitly as
 \begin{equation}
 \mathlarger{\tilde e}_{\mathsmaller{\Delta}_{\alpha}}(k)=\frac{2}{\sqrt{2\pi}}\frac{1}{k}e^{-{\rm i} k\bar x_\alpha}\sin\left(\frac{k\delta}{2}\right).
 \end{equation}
 Next we introduce the action~$S_{Pk}$
 \begin{equation}
S_{Pk}=S_P+\hbar k\bar x,
\end{equation}
where~$S_{P}$ is defined in Eq.~\eqref{S_P}. 
Then the sum over particle paths in Eq.~\eqref{particle-path-int1} becomes
\begin{align}\label{particle-path-int2}
\int_{c_{\alpha}}&\delta x(t)\,{\rm e}^{\frac{\rm i}{\hbar} S_P[x(t)]}=\nonumber\\
&
\frac{1}{\pi}\int_{-\infty}^\infty\frac{{\rm d}k}{k}e^{-ik\bar x_{\alpha}}\sin \left (\frac{k\delta}{2}\right)\nonumber\\
&\times
\mathlarger{\int}\limits_{(x',0)}^{(x,T)}\delta x(t){\rm e}^{\frac{\rm i}{\hbar}S_{Pk}[x(t)]},
\end{align}
where we have introduced an effective action
\begin{align}\label{S_Pk}
S_{Pk}[x(t)]=&\mathlarger{\int}_0^T{\rm d}t\,\Big (\,\frac{m}{2}\dot x^2-V(x,t)\nonumber\\
&
-\frac{f(t)}{T}xP+\frac{f(t)}{T}\hbar kx \,\Big ).
\end{align}

When we insert Eq.~\eqref{particle-path-int2} into Eq.~\eqref{particle-path-int1} and the resulting expression into Eq.~\eqref{class -operator1}, for the matrix elements of the class operators we obtain 
\begin{widetext}
\begin{equation}\label{class-operator3}
\left\langle {x,X} \right|\widehat C_{\alpha}\left| {x',X'} \right\rangle =
 \int_{ - \infty }^\infty  \frac{dP}{2\pi\hbar } e^{ \frac{\rm i}{\hbar}[P(X - X') -(P^2/2M)T]}
\frac{1}{\pi}\int_{-\infty}^\infty\frac{{\rm d}k}{k}e^{-ik\bar x_{\alpha}}\sin \left (\frac{k\delta}{2}\right)
\mathlarger{\int}\limits_{(x',0)}^{(x,T)}\delta x(t){\rm e}^{\frac{\rm i}{\hbar}S_{Pk},[x(t)]},
\end{equation}
\end{widetext}
where the effective action~$S_{Pk}$ is given by Eq.~\eqref{S_Pk}.  This expression can be applied to any potential and any coupling function.  

We can rewrite Eq.~\eqref{class -operator1} as
\begin{align}\label{CLASS-OP5}
\left\langle {x,X} \right|\widehat C_{\alpha}\left| {x',X'} \right\rangle&=
\!\! \int_{ - \infty }^\infty \!\frac{dP}{2\pi\hbar } e^{ \frac{\rm i}{\hbar}[P(X - X') -(P^2/2M)T]}\nonumber\\
 &\times
\left\langle {x} \right|\widehat C_{P\alpha}\left| {x'} \right\rangle,
\end{align}
where~$\left\langle {x} \right|\widehat C_{P\alpha}\left| {x'} \right\rangle$ is the matrix element of a reduced class operator which is defined by Eq.~\eqref{particle-path-int2}.

 In the present work we will restrict ourselves to the two tractable cases: a free particle, and a harmonic oscillator.  In both cases the coupling function will be set to be constant in the interval~$[0,T]$.

\section{The Free particle}\label{free particle}
\subsection{The class operators}

For a free particle~$V(x,t)=0$, and the only interaction is with the apparatus.  In addition~$f(t)=g$, a dimensionless coupling constant between the particle and the apparatus. In this case since~$f(t)$ is a constant, it will be convenient to use
\begin{equation}\label{position-average}
\bar x=\frac{1}{T}\int_0^T\!\!\!x(t){\rm d}t
\end{equation}
 instead of Eq.~\eqref{weighted-bar x} in the definition~\eqref{histories} for the mutually exclusive coarse-grained classes.
The pointer coordinate will correlate with~$g\bar x$.

 Now the effective action in Eq.~\eqref{S_Pk} is replaced by the effective action corresponding to taking the average according to Eq.~\eqref{position-average} instead of Eq.~\eqref{weighted-bar x}. That is,
\begin{equation}\label{S_Pkfree}
S_{Pk}=\mathlarger{\int}_0^T{\rm d}t\left[ \frac{m}{2}\dot x^2+\frac{1}{T}(k\hbar-gP)x  \right].
\end{equation}

Using a well-known result in Ref.~\cite{Feynman}, the sum over particle paths in the expression~\eqref{class-operator3} can be carried out to obtain 
\begin{align}\label{sum_part_paths}
 &\mathlarger{\int}\limits_{(x',0)}^{(x,T)}\delta x(t){\rm e}^{\frac{\rm i}{\hbar}S_{Pk},[x(t)]}=
\left(\frac{m}{2\pi{\rm i}\hbar T}\right)^{\!\!1/2}\!\!\!\!{\rm exp}\Bigg\{\frac{{\rm i}}{\hbar}\bigg [  \frac{m}{2T}(x-x')^2\nonumber\\
&+
\frac{1}{2}(\hbar k-gP)(x+x')-(\hbar k-gP)^2\frac{T}{24m}   \bigg ]     
 \Bigg\}.
\end{align}
Inserting Eq.~\eqref{sum_part_paths} into Eq.~\eqref{particle-path-int2} for the matrix elements of the reduced class operators we obtain 
\begin{widetext}
\begin{align}\label{reduced-class-op}
\left\langle {x} \right|\widehat C_{P\alpha}\left| {x'} \right\rangle =
 &\left(\frac{m}{2\pi{\rm i}\hbar T}\right)^{1/2}{\rm exp}\Bigg\{\frac{\rm i}{\hbar}\bigg [\frac{m}{2T}(x-x')^2-\frac{1}{2}gP(x+x')-g^2P^2\frac{T}{24m}\bigg]\Bigg\}\nonumber\\
 &\times
\frac{1}{\pi}{\mathlarger\int_{-\infty}^\infty}\frac{{\rm d}k}{k}{\rm e}^{-ik\bar x_{\alpha}}\sin \left (\frac{k\delta}{2}\right)
{\rm exp}\Bigg\{{\rm i}\left[k\left(\frac{x+x'}{2}+gP\frac{T}{12m} \right)-\hbar k^2\frac{T}{24m}\right] \Bigg\}.
\end{align}
\end{widetext}

Next we insert the expression~\eqref{reduced-class-op} into Eq.~\eqref{class-operator3}. Carrying out the integration over the momentum of the pointer we obtain the result
\begin{align}\label{class-operator4}
&\left\langle {x,X} \right|\widehat C_{\alpha}\left| {x',X'} \right\rangle =\left\langle {x, X} \right|{\rm e}^{-\frac{\rm i}{\hbar}\hat HT}\left| {x'.X'} \right\rangle\nonumber\\
&\times
\frac{1}{\pi}{\mathlarger\int_{-\infty}^{\infty}}\frac{{\rm d}k}{k}{\rm e}^{-ik\bar x_{\alpha}}\sin \left (\frac{k\delta}{2}\right){\rm e}^{\frac{\rm i}{\hbar}(-\ell^2k^2+Zk)},
\end{align}
where~$\hat H$ is the Hamiltonian of the system given by Eq.~\eqref{H} for a free particle and a constant coupling function.  In this case the propagator for the system is
\begin{align}\label{TOT-prop}
\left\langle {x, X} \right|&{\rm e}^{-\frac{\rm i}{\hbar}\hat HT}\left| {x'.X'} \right\rangle=\frac{(mM_{\it eff})^{1/2}}{2\pi{\rm i}\hbar T}{\rm exp}\bigg\{\frac{\rm i}{\hbar}\frac{m}{2T}(x-x')^2\bigg\}\nonumber\\
&\times
{\rm exp}\Bigg\{\frac{M_{\it eff}}{2T}\left [ X-X'-g\frac{(x+x')}{2}\right ]^2\Bigg\},
\end{align}
with the effective mass
\begin{equation}\label{eff-mass}
M_{\it eff}=M\left(1+\frac{g^2M}{12m}\right)^{-1}.
\end{equation}
The results, Eq.~\eqref{TOT-prop} and Eq.~\eqref{eff-mass}, were already obtained in Refs.~\cite{FR-1,FR}.  The expressions for the lengths~$\ell$ and~$Z$ in the exponential in the integrantion over~$k$ are given by
\begin{equation}\label{ell}
\ell=\left(\frac{\hbar T}{6m}\frac{M_{\it eff}}{M}\right)^{1/2}
\end{equation}
and
\begin{equation}\label{Z}
Z=\frac{gM_{\it eff}}{12m}\left[X-X'-g(x+x')/2\right]+(x+x')/2.
\end{equation}
The integration over~$k$ in Eq.~\eqref{class-operator4} yields the result for the matrix elements of the class operators 
\begin{equation}\label{exact-class-op}
\!\!\!\!\!\!\!\!\!\left\langle {x,X} \right|\widehat C_{\alpha}\left| {x',X'} \right\rangle \!=\! \left\langle {x, X} \right|{\rm e}^{-\frac{\rm i}{\hbar}\hat HT}\left| {x',X'} \right\rangle 
\!{\mathlarger E}_{\!\Delta_{\alpha}}\!\!\left(Z,\ell\right),
\end{equation}
where we have introduced the function of the two lengths~$\ell$ and~$Z$
\begin{align}\label{define-E}
{\mathlarger E}_{\Delta_{\alpha}}\left(Z,\ell\right)=&
\frac{1}{2}\Bigg\{{\rm erf}\left[\frac{1}{\sqrt{\rm i}\ell}(Z-\bar x_{\alpha}
+\delta/2)\right]\nonumber\\
&
-{\rm erf}\left[\frac{1}{\sqrt{\rm i}\ell}(Z-\bar x_{\alpha}-\delta/2)\right]\Bigg\}.
\end{align}

\subsection{The classical limit of the class operators}

We now evaluate the matrix elements of the class operators in the classical limit,~$\hbar\rightarrow 0$.  
Recall
\begin{equation}\label{tot-class-op}
\!\!\left\langle {x,X} \right|\widehat C_{\alpha}\left| {x',X'} \right\rangle =
\!\!\int_U\!\int_u\!\!\delta Y\delta x\,\mathlarger{e}_{\mathsmaller{\Delta}_{\alpha}}\left(\bar x\right){\rm e}^{\frac{\rm i}{\hbar}S[x(t),X(t)]},
\end{equation}
where~$S$ is the action in Eq.~\eqref{S1} with~$f(t)=g$ and~$V(x,t)=0$.

We introduce the change of path variables
\begin{equation}\label{x-variable-change}
x(t)=x_{cl}(t)+y(t)
\end{equation}
\begin{equation}\label{X-variable-change}
 X(t)=X_{cl}(t)+Y(t),
\end{equation}
where~$x_{cl}(t)$ and~$X_{cl}(t)$ are the classical paths for the particle and for the pointer respectively.  Thus~$x_{cl}(t)$ and~$X_{cl}(t)$ satisfy the classical equations of motion for the action~$S$ with~$x_{cl}(0)=x'$,~$x_{cl}(T)=x$ and~$X_{cl}(0)=X'$,~$X_{cl}(T)=X$. Substituting the changes~\eqref{x-variable-change} and~\eqref{X-variable-change} into Eq.~\eqref{tot-class-op}
we obtain
\begin{align}\label{y-Y-int}
&\left\langle {x,X}\right|\widehat C_{\alpha}\left| {x',X'} \right\rangle =\iint\delta Y(t)\delta y(t)\nonumber\\
&\times
\mathlarger{e}_{\mathsmaller{\Delta}_{\alpha}}\left(\bar x[x_{cl}(t)+y(t)]\right){\rm e}^{\frac{\rm i}{\hbar}S[x_{cl}(t),X_{cl}(t)]}\nonumber\\
&\qquad\times
{\rm exp}\Bigg[ \frac{\rm i}{\hbar}\int_0^T{\rm d}t\Big(\frac{m}{2}\dot y^2+\frac{M}{2}\dot Y^2\nonumber\\
&\qquad\quad\quad
+\frac{gM}{T}y\dot Y+\frac{gM}{2T^2}y^2 \Big) \Bigg].
\end{align}
In the~$\hbar\rightarrow 0$ limit the exponent in the path integral over~$Y(t)$ and~$y(t)$ becomes very large, and the path integral is dominated by the stationary path at the saddle point~$y(t)=0$,~$Y(t)=0$.  In this limit the function~$\mathlarger{e}_{\mathsmaller{\Delta}_{\alpha}}$ in Eq.~\eqref{y-Y-int} can be evaluated at the saddle point and can be pulled out of the integral to yield
\begin{align}\label{cl-limit1}
\!\!\!\!\left\langle {x,X}\right|\widehat C_{\alpha}\left| {x',X'} \right\rangle \approx \mathlarger{e}_{\mathsmaller{\Delta}_{\alpha}}\left(\bar x_{cl}\right)
\left\langle {x, X} \right|{\rm e}^{-\frac{\rm i}{\hbar}HT}\left| {x'.X'} \right\rangle
\end{align}
where the expression~\eqref{2D-pathint} for the propagator of the system has been re-expressed in terms of a path integral over~$y(t)$ and~$Y(t)$. That is, 
\begin{align}\label{TOT-prop2}
\left\langle {x, X} \right|&{\rm e}^{-\frac{\rm i}{\hbar}\hat HT}\left| {x'.X'} \right\rangle=\nonumber\\
&
{\rm e}^{\frac{\rm i}{\hbar}S[x_{cl}(t),X_{cl}(t)]}\iint\delta Y(t)\delta y(t)\nonumber\\
&\qquad\times
{\rm exp}\Bigg[ \frac{\rm i}{\hbar}\int_0^T{\rm d}t\Big(\frac{m}{2}\dot y^2+\frac{M}{2}\dot Y^2\nonumber\\
&\qquad\quad\quad
+\frac{gM}{T}y\dot Y+\frac{gM}{2T^2}y^2 \Big) \Bigg].
\end{align}

The classical equations of motion for this system follow from the action~\eqref{S1} with~$f(t)=g$ and~$V(x,t)=0$
\begin{equation}
m\ddot x_{cl}+\frac{gM}{T}\dot X_{cl}-\frac{g^2M}{T^2}x_{cl}=0
\end{equation}
\begin{equation}
\frac{{\rm d}}{{\rm d}t}\left [M\left(\dot X_{cl}-\frac{gx_{cl}}{T}\right)\right]=0.
\end{equation}
The solution with~$x_{cl}(0)=x' , \,x_{cl}(T)=x$ and~$X_{cl}(0)=X' ,\, X_{cl}(T)=X$ is
\begin{align}\label{class-path}
x_{cl}(t)&=x'+\left(x-x'+\frac{gM}{2m}X_{0T}\right)\frac{t}{T}\nonumber\\
&
-\frac{gM}{2m}X_{0T}\frac{t^2}{T^2}
\end{align}
\begin{equation}
\dot X_{cl}(t)=\frac{g}{T}x_{cl}(t)+\frac{X_{0T}}{T},
\end{equation}
where
\begin{equation}
X_{0T}=\left[X-X'-g(x+x')/2\right]\frac{M_{\it eff}}{M}
\end{equation}
and~$M_{ef\!f}$ is defined in Eq.~\eqref{eff-mass}.
The action evaluated at the classical path is thus
\begin{align}\label{S_cl}
S_{cl}&=\frac{1}{T}\Bigg\{ \frac{m}{2}(x-x')^2 \nonumber\\
&
+\frac{M_{\it eff}}{2}\left[ X-X'-g\frac{(x+x')}{2}\right]^2
\Bigg\},
\end{align}
which, when inserted into Eq.~\eqref{TOT-prop2}, yields the expression~\eqref{TOT-prop} for the propagator of the system.  The amplitude factor for the propagator is obtained from the path integral over~$y(t)$ and~$Y(t)$.  That is,
\begin{align}\label{x-cl}
\iint&\delta Y(t)\delta y(t){\rm exp}\Bigg[ \frac{\rm i}{\hbar}\int_0^T{\rm d}t\Big(\frac{m}{2}\dot y^2+\frac{M}{2}\dot Y^2\nonumber\\
&
+\frac{gM}{T}y\dot Y+\frac{gM}{2T^2}y^2 \Big) \Bigg]=\frac{(mM_{\it eff})^{1/2}}{2\pi{\rm i}\hbar T}.
\end{align}

From the expression~\eqref{class-path} we obtain the time average of the classical path during time~$T$
\begin{equation}\label{bar x-cl}
\!\!\bar x_{cl}=\frac{x+x'}{2}+\frac{g}{12m}M_{\it eff}\left[X-X'-g\left( \frac{x+x'}{2}\right) \right].
\end{equation}
Then in Eq.~\eqref{cl-limit1} we obtain the classical limit for the matrix elements of the class operators
\begin{align}\label{classical-lim}
&\left\langle {x,X}\right|\widehat C_{\alpha}\left| {x',X'} \right\rangle \approx\mathlarger{e}_{\mathsmaller{\Delta}_{\alpha}}\Bigg(\frac{x+x'}{2}+\nonumber\\ 
&
\frac{g}{12m}M_{\it eff}\left[X-X'-g\left( \frac{x+x'}{2}\right) \right]\Bigg)\nonumber\\
&\times
\left\langle {x, X} \right|{\rm e}^{-\frac{\rm i}{\hbar}HT}\left| {x'.X'} \right\rangle.
\end{align}
Comparing the argument of the function~$\mathlarger{e}_{\mathsmaller{\Delta}_{\alpha}}$ in the expression ~\eqref{classical-lim} with the expression for~$Z$ in the definition~\eqref{Z}, we identify~$Z=\bar x_{cl}$.

Next, as observed in Ref.~\cite{B-H},~${\mathlarger E}_{\Delta_{\alpha}}\left(Z,\ell\right)$ approximates~$\mathlarger{e}_{\mathsmaller{\Delta}_{\alpha}}\left(Z/\ell\right)$ as~$\hbar\rightarrow 0$.  Therefore  in the classical limit the exact expression~\eqref{exact-class-op} for the matrix elements of the class operators yields the result~\eqref{classical-lim}. 

Let us examine how~$\widehat C_{\alpha}$ acts on a wavefunction. 
If the initial wavefunction consists of a wavepacket for the particle, and the pointer is in the sharp position wavefunction~$\delta(X)$, then the branch wavefunction is 
\begin{align}
\!\Psi_{\alpha}(x,X,T)=\!\!\int_{-\infty}^\infty\!\!\!&{\rm d}x'K(x,X,T;x',0,0)\nonumber\\
&\times
{\mathlarger E}_{\Delta_{\alpha}}\left(Z,\ell\right)\varphi_0(x'),
\end{align}
where~$K(x,X,T;x',X',0)$ is the propagator of the system defined by the unrestricted sum over paths~\cf Eq.~\eqref{2D-pathint}, and~${\mathlarger E}_{\Delta_{\alpha}}\left(Z,\ell\right)$ is defined by the expression~\eqref{define-E}.

In the classical limit this expression becomes
\begin{align}
\!\Psi_{\alpha}(x,X,T)\approx\!\!\int_{-\infty}^\infty\!\!\!&{\rm d}x'K(x,X,T;x',0,0)\nonumber\\
&\times
{\mathlarger e}_{\Delta_{\alpha}}\left(\bar x_{cl}\right)\varphi_0(x'),
\end{align}
which shows that in the classical limit the branch wavefunction is the result of the time evolution of the wavepacket
\begin{equation}
\varphi_\alpha(x)={\mathlarger e}_{\Delta_{\alpha}}\left(\bar x_{cl}\right)\varphi_0(x).
\end{equation}

Consider the branch wavefunction
with~$\left| {\Psi_0} \right\rangle=\left| {\varphi_0} \right\rangle\left| {\Phi_0} \right\rangle$,
where~$\left| {\varphi_0} \right\rangle$
has a gaussian wavefunction of width~$d$ describing the initial state of the particle
\begin{equation}
\varphi_0(x)=\left( \frac{2}{\pi d^2} \right)^{1/4}{\rm exp}\left(-\frac {x^2 }{ d^2}\right),
\end{equation}
and~$ \left| {\Phi_0} \right\rangle$ has a wavefunction~$\Phi_0(X) $ that describes the initial state of the pointer, typically a wavepacket of some width~$D$.
With this initial state we expect classical behavior for spacetime alternatives when (1) the coarse-graining parameter ~$\delta>>d$, and (2) the duration of the spacetime alternative for this system is such that~$T>>t_{\rm spread}$, where~$t_{\rm spread}$ is the wavepacket spreading time if we treat the particle as a free particle with negligible interaction with the pointer and
\begin{equation}
t_{\rm spread}=\frac{d^2 m}{2\hbar}.
\end{equation}

In Ref.~\cite{FR} it was argued that for a narrow wavepacket describing the initial state of the particle, the entanglement between particle and pointer can be neglected for measurement durations which are less than~$t_{\rm spread}$.  That is, the particle evolves freely in this approximation.  With this in mind let~$d<<\delta$ for a very narrow wavepacket and for~$T<<t_{\rm spread}$, so that from Eq.~\eqref{ell}
\begin{equation}
T=\frac{6m\ell^2}{\hbar}\frac{M}{M_{\it eff}}
\end{equation}
and
\begin{equation}
\frac{d}{\ell}>>\sqrt{\frac{6M}{M_{\it eff}}}>1.
\end{equation}
Thus the conditions (1) and (2) for classical behavior are expressed as
\begin{equation}
\frac{\delta}{\ell}>>\frac{d}{\ell}>>1.
\end{equation}
This is the condition already verified in Ref.~\cite{B-H} to ensure that in the limit~$\hbar\rightarrow 0$
\begin{equation}
{\mathlarger E}_{\Delta_{\alpha}}\left(Z,\ell\right)\approx \mathlarger{e}_{\mathsmaller{\Delta}_{\alpha}}(Z/\ell).
\end{equation}


\subsection{The decoherence functional.  Decoherence in the classical limit}
We will next show explicitly  the decoherence in the classical limit of coarse-grained spacetime alternatives defined in~\eqref{histories} .
First we consider the matrix elements of the decoherence functional in the position representation
 \begin{align}\label{decoher-funct}
 \left\langle{x,X}\right|\widehat C_{\alpha}^{\dagger}\widehat C_{\alpha'}\left| {x',X'} \right\rangle=\iint
 &{\rm d}y{\rm d}Y\left\langle {y,Y}\right|\widehat C_{\alpha}\left| {x,X} \right\rangle^*\nonumber\\
\!\!\!\!\!\times&
\left\langle {y,Y}\right|\widehat C_{\alpha'}\left| {x',X'} \right\rangle.
 \end{align}
 Next, using Eq.~\eqref{S_cl} we can rewrite Eq.~\eqref{TOT-prop} as
 \begin{equation}
\!\!\!\!\!\left\langle {x, X} \right|{\rm e}^{-\frac{\rm i}{\hbar}HT}\left| {x'.X'} \right\rangle\!=\!\frac{(mM_{\it eff})^{1/2}}{2\pi{\rm i}\hbar T}{\rm e}^{\frac{\rm i}{\hbar}S_{cl}(x,X,x',X')}.
 \end{equation}
 This expression for the propagator of the system can then be introduced into the expression~\eqref{exact-class-op} for the exact matrix elements of the class operators, and the result can be introduced into Eq.~\eqref{decoher-funct}. We obtain 
 \begin{align}\label{decoher-funct1}
 &\left\langle{x,X}\right|\widehat C_{\alpha}^{\dagger}\widehat C_{\alpha'}\left| {x',X'} \right\rangle=
 \frac{mM_{\it eff}}{(2\pi\hbar T)^2}\int_{-\infty}^\infty\int_{-\infty}^\infty{\rm d}y\,{\rm d}Y\nonumber\\
 &\times
 {\rm exp}\left\{\frac{{\rm i}}{\hbar}\left\{S_{cl}(x',X';y,Y)-S_{cl}(x,X;y,Y)\right\}\right\}\nonumber\\
 &\times
  {\mathlarger E}^*_{\Delta_{\alpha}}(\bar x_{cl},\ell)
{\mathlarger E}_{\Delta_{\alpha'}}\!(\bar x'_{cl},\ell),
\end{align}
where
\begin{equation}
\bar x_{cl}=\bar x [x_{cl}(t,y,x,Y,X)]
\end{equation}
and
\par\vspace{-4ex}
\begin{equation}
 \bar x'_{cl}=\bar x [x_{cl}(t,y,x',Y,X'],
\end{equation}
with~$\bar x_{cl}$ having the form shown by the expression~\eqref{bar x-cl}.  

The explicit integral form for the matrix elements of the decoherence functional is then 
\begin{widetext}
\begin{align}\label{DECOHERENCE-functional}
 &\left\langle{x,X}\right|\widehat C_{\alpha}^{\dagger}\widehat C_{\alpha'}\left| {x',X'} \right\rangle=\frac{mM_{\it eff}}{(2\pi\hbar T)^2}{\rm exp}\left[ \frac{{\rm i}m}{2\hbar T}(x'^2-x^2)\right] 
 {\rm exp}\left\{\frac{{\rm i}M_{\it eff}}{2\hbar T}\left[ X'^2-X^2+g\left( X'x'-Xx\right)
 +
 \frac{g^2}{4}(x'^2-x^2)\right] \right\}\nonumber\\
 &\qquad\qquad\quad\times
 \int_{-\infty}^\infty\int_{-\infty}^\infty{\rm d}y\,{\rm d}Y{\rm exp}\left[  \frac{{\rm i}m}{\hbar T}y(x-x')\right]{\rm exp}\left\{\frac{{\rm i}M_{\it eff}}{2\hbar T}\!\left[2Y(X-X')+g(x-x')Y-gy(X-X')  \right] \right\}\nonumber\\
 &\times
 {\mathlarger E}^*_{\Delta_{\alpha}}\left(\frac{gM_{\it eff}}{12m}\left[Y-X-g(y+x)/2\right]+(y+x)/2,\ell\right)
{\mathlarger E}_{\Delta_{\alpha'}}\left(\frac{gM_{\it eff}}{12m}\left[Y-X'-g(y+x')/2\right]+(y+x')/2,\ell\right).
\end{align}
\end{widetext}
The matrix elements described by Eq.~\eqref{DECOHERENCE-functional} do not lead to decoherence, even for the case of a particle in a sharp position state at~$t=0$. That is,
\begin{equation}
\Braket{\Phi_0,x_0|\widehat C_{\alpha'}^{\dagger}\,\,\widehat C_\alpha|\Phi_0,x_0}\neq 0,
\end{equation}
where~$\Phi_0(X)$ is the initial wavefunction for the pointer.

In the classical limit~$\hbar\rightarrow 0$ the length~$\ell$ introduced in the definition~\eqref{ell} becomes small  and, as discussed in Ref.~\cite{B-H}, we obtain the limit
\begin{equation}
{\mathlarger E}_{\Delta_{\alpha}}\left(Z,\ell\right)\xrightarrow[\hbar \rightarrow 0] {}\mathlarger{e}_{\mathsmaller{\Delta}_{\alpha}}(Z/\ell).
\end{equation}
This allows the~$E$'s in Eq.~\eqref{decoher-funct1} to be replaced by the~$e$'s. The~$e$'s are to be evaluated at the classical paths for the particle and the pointer, as follows from the expression~\eqref{cl-limit1} for the matrix of the  classical limit of the class operators and from Eq.~\eqref{decoher-funct} for the matrix elements of the decoherence functional. Thus in the integration in Eq.~\eqref{DECOHERENCE-functional} we make the replacements
\begin{align}
&{\mathlarger E}_{\Delta_{\alpha}}\!\!\left(\frac{gM_{\it eff}}{12m}\left[Y-X-g(y+x)/2\right]+(y+x)/2,\ell\right)\nonumber\\
&
\rightarrow \mathlarger{e}_{\mathsmaller{\Delta}_{\alpha}}(\bar x_{cl}(t,y,x,Y,X)),\\
&
{\mathlarger E}_{\Delta_{\alpha'}}\!\!\left(\frac{gM_{\it eff}}{12m}\left[Y-X'-g(y+x')/2\right]+(y+x')/2,\ell\right)\nonumber\\
&
\rightarrow \mathlarger{e}_{\mathsmaller{\Delta}_{\alpha'}}(\bar x_{cl}(t,y,x',Y,X')).
\end{align}
The classical path in~$e_{\alpha}$ for the particle starts at~$x$ and ends at~$y$ at time~$T$, and for the pointer it starts at~$X$ and ends at~$Y$ at time~$T$.  For~$e_{\alpha'}$ the path for the particle starts at~$x'$ and ends at~$y$ at time~$T$, and for the pointer it starts at~$X'$ and ends at~$Y$ at time~$T$.
In the~$\hbar\rightarrow 0$ limit we can thus rewrite Eq.~\eqref{decoher-funct1}
\begin{align}\label{class-lim}
&\left\langle{x,X}\right|\widehat C_{\alpha}^{\dagger}\widehat C_{\alpha'}\left| {x',X'} \right\rangle\sim\iint{\rm d}y\,{\rm d}Y\nonumber\\
&\times
\mathlarger{e}_{\mathsmaller{\Delta}_{\alpha}}\left(\bar x_{cl}(t,y,x,Y,X)\right)\mathlarger{e}_{\mathsmaller{\Delta}_{\alpha}}\left(\bar x_{cl}(t,y,x',Y,X')\right)\nonumber\\
&\times
{\rm exp}\left\{\frac{{\rm i}}{\hbar}\left\{S_{cl}(x',X';y,Y)-S_{cl}(x,X;y,Y)\right\}\right\}.
\end{align}
\par\vspace{1ex}
The exponent in the integrand vanishes if~$x=x'$ and~$X=X'$,  and so does the factor multiplying the exponential because of the exclusive ranges for the~$e$'s.  By the Riemann-Legesgue lemma~\cite{Bender} the integral vanishes in the limit~$\hbar\rightarrow 0$.  This confirms Eq.~\eqref{decoherence} for decoherence in the classical limit.

\section{The Harmonic Oscillator}

For the oscillator we set~$V(x,t)=\frac{m}{2}\omega^2 x^2$ in the Hamiltonian~\eqref{H}, and in addition~$f(t)=g$, a dimensionless constant, in the interaction~\eqref{Hi} between the oscillator and the apparatus. 
In the expression~\eqref{CLASS-OP5} for the matrix elements of the class operators of the system, the matrix elements~$\left\langle {x} \right|\widehat C_{P\alpha}\left| {x'} \right\rangle$ of the reduced class operators   are given by the expression~\eqref{particle-path-int2},
 where the action~$S_{Pk}$ is now
\begin{equation}\label{S_PkSHO}
S_{Pk}=\mathlarger{\int}_0^T{\rm d}t\left[ \frac{m}{2}\dot x^2-\frac{m}{2}\omega^2 x^2+\frac{1}{T}(k\hbar-gP)x  \right].
 \end{equation}\\
 
Next ,using another well-known result from Ref.~\cite{Feynman}, we can write the path integral for the system defined by the action~\eqref{S_PkSHO} as
\par\vspace{-3ex}
\begin{equation}
\int_u\delta x(t){\rm e}^{\frac{{\rm i}}{\hbar}S_{Pk}[x(t)]}=A{\rm e}^{\frac{{\rm i}}{\hbar}S_{Pk,cl}},
\end{equation}\\
with the amplitude factor
\begin{equation}
A=\left(\frac{m\omega}{2\pi{\rm i}\hbar\sin\omega T}\right)^{1/2}.
\end{equation}
The value of the classical action is
\begin{align}\label{S_Pk,cl}
&S_{Pk,cl}=\frac{m\omega}{2\sin\omega T}\Bigg[(x^2+x'^2)\cos\omega T-2xx'\nonumber\\
&+\frac{2}{m\omega T}(\hbar k-gP)\frac{(1-\cos\omega T)}{\omega}(x+x')\nonumber\\
&-\frac{2}{m^2\omega^2 T^2}(\hbar k-gP)^2\left( \frac{1-\cos\omega T}{\omega^2}-\frac{T}{2\omega}\sin\omega T  \right)
\Bigg].
\end{align}

The expression for the matrix elements for the reduced class operators in Eq.~\eqref{particle-path-int2} is
\begin{equation}
\!\!\left\langle {x} \right|\widehat C_{P\alpha}\left| {x'} \right\rangle=\frac{A}{\pi}\int_{-\infty}^\infty\frac{{\rm d}k}{k}{\rm e}^{-{\rm i}k\bar x_{\alpha}}\sin\left (\frac{k\delta}{2}  \right ){\rm e}^{\frac{{\rm i}}{\hbar}S_{Pk,cl}},
\end{equation}
where~$S_{Pk,cl}$ is given by the Eq.~\eqref{S_Pk,cl} above.

The expression for the matrix elements of the class operators for the system is then
\begin{align}
&\left\langle {x,X} \right|\widehat C_{\alpha}\left| {x',X'} \right\rangle =\frac{A}{\pi}\int_{-\infty}^\infty\frac{{\rm d}k}{k}{\rm e}^{-{\rm i}k\bar x_{\alpha}}\sin\left (\frac{k\delta}{2} \right )\nonumber\\\nonumber\\
&\qquad\qquad\times
\int_{-\infty}^\infty\frac{{\rm d}P}{2\pi\hbar}
{\rm e}^{-\frac{{\rm i}}{\hbar}\frac{P^2}{2M}T+  \frac{{\rm i}}{\hbar}P(X-X') }{\rm e}^{\frac{{\rm i}}{\hbar}S_{Pk,cl}}.
\end{align}

The integration over the momentum~$P$ of the pointer can be carried out to obtain the result
\begin{widetext}
\begin{align}
&\left\langle {x,X} \right|\widehat C_{\alpha}\left| {x',X'} \right\rangle =\left\langle {x,X} \right|{\rm e}^{-\frac{{\rm i}}{\hbar}HT} \left| {x',X'} \right\rangle\frac{1}{\pi}\int_{-\infty}^\infty\frac{{\rm d}k}{k}{\rm e}^{-{\rm i}k\bar x_{\alpha}}\sin\left (\frac{k\delta}{2} \right )\nonumber\\
&\qquad\qquad\times
{\exp}\Bigg\{{\rm i}\left [\frac{g}{\omega^2 T^2}\frac{M_{\it eff}}{m}\left( X-X'-g(T)\frac{x+x'}{2} \right)\left(\frac{g(T)}{g}-1 \right)+\left( \frac{1-\cos\omega T}{\omega T\sin\omega T}\right)(x+x') \right ]k
\nonumber\\ 
&\qquad\qquad\qquad\qquad\qquad\qquad\qquad
+\frac{{\rm i\hbar}T}{m}\left [ \frac{g^2}{2\omega^4T^4}\frac{M_{\it eff}}{m}\left(\frac{g(T)}{g}-1 \right)^2
-\frac{1}{2\omega^2T^2}
\left(\frac{g(T)}{g}-1 \right) \right]k^2
\Bigg\},
\end{align}
\end{widetext}
where the propagator for the system was obtained in Ref.~\cite{FR-1}.  
The effective mass~$M_{ef\!f}$ was given by
\begin{equation}
M_{\it eff}=\frac{M}{1+\mathlarger{\frac{g^2M}{\omega^2T^2m}\left[\frac{g(T)}{g}-1 \right]}},
\end{equation}
with the time-dependent dimensionless coupling constant
\begin{equation}\label{g(T)}
g(T)=\frac{2g}{\omega T}\tan\left(\frac{\omega T}{2}\right).
\end{equation}

Finally, the integration over~$k$ can be carried out to yield the exact expression for the matrix elements of the class operators
\begin{equation}\label{exact-class-opSHO}
\!\!\!\!\!\!\!\left\langle {x,X} \right|\widehat C_{\alpha}\left| {x',X'} \right\rangle = \left\langle {x, X} \right|{\rm e}^{-\frac{\rm i}{\hbar}\hat HT}\left| {x'.X'} \right\rangle 
\!{\mathlarger E}_{\Delta_{\alpha}}\!\left(Z,\ell\right).
\end{equation}
This is of the same form as~\eqref{exact-class-op} with
\begin{equation} \label{H_SHO}
\hat H=\frac{\hat p^2}{2m}+\frac{m}{2}\omega^2 x^2+\frac{\hat P^2}{2M}+ \frac{g}{T}  x\hat P.
\end{equation}
The function~$E_{\Delta_{\alpha}}$ is defined in~\eqref{define-E}.  For the harmonic oscillator the length~$Z$ is given by the expression
\begin{align}\label{Z-SHO}
Z=&\left(\frac{1-\cos\omega T}{\omega T\sin\omega T}\right)(x+x')+\frac{g }{\omega^2T^2 }\frac{M_{\it eff}}{m}\bigg[ X-X'\nonumber\\
&-g(T)\left(\frac{x+x'}{2} \right)\bigg]\left(\frac{g(T)}{g}-1 \right),
\end{align}
and the length~$\ell$ is replaced by
\begin{align}\label{l-SHO}
\ell=\left[\frac{2\hbar }{m\omega^2T}\left(  \frac{g(T)}{g}-1\right)\frac{M_{\it eff}}{M}  \right]^{1/2}.
\end{align}

Next, before taking the classical limit, we will evaluate the classical value for the average position of a particle.  To this effect we will evaluate the classical path from the action
\begin{align}\label{S-SHO}
S\left[ {x(t),X(t)} \right] = &\mathlarger{{\mathlarger{\int}}}_{\!\!\!\!\!0}^T\Bigg[ \frac{m}{2}\dot x^2  - \frac{m}{2}\omega^2x^2\nonumber\\
&  
 +\frac{M}{2}\left( {\dot X - \frac{g}{T}x} \right)^2\Bigg] {\rm d}t.
\end{align}
From Eq.~\eqref{S-SHO} we obtain the classical equations of motion 
\begin{equation}
m\ddot x_{cl}+\frac{gM}{T}\dot X_{cl}-\frac{g^2M}{T^2}x_{cl}+m\omega^2x_{cl}=0,
\end{equation}
and
\begin{equation}
\frac{{\rm d}}{{\rm d}t}\left [M\left(\dot X_{cl}-\frac{gx_{cl}}{T}\right)\right]=0.
\end{equation}

 With the end point conditions~$x_{cl}(0)=x'$,~$x_{cl}(T)=x$ 
and~$X_{cl}(0)=X'$, ~$X_{cl}(T)=X$
these equations are solved for the position of the particle
\begin{widetext}
\begin{align}\label{cl-particle}
x_{cl}(t)=&\left\{x'+\frac{gM_{\it eff}}{\omega^2T^2m}\left[ X-X'-g(T)\left(\frac{x+x')}{2} \right) \right]  \right\}\cos\omega t\nonumber\\
&+
\frac{1}{\sin\omega T}\left\{x-x'\cos\omega T-(\cos\omega T-1)\frac{gM_{\it eff}}{\omega^2T^2m}\left[ X-X'-g(T)\left(\frac{ x+x'}{2}\right)\right]  \right\}\sin\omega t\nonumber\\
&-
\frac{gM_{\it eff}}{\omega^2T^2m}\left[X-X'-g(T)\left( \frac{x+x'}{2}\right) \right],
\end{align}
\end{widetext}
and for the velocity of the pointer
\begin{widetext}
\begin{align}\label{cl-pointer}
\dot X_{cl}(t)=&\frac{g}{T}\left\{ x'+\frac{gM_{\it eff}}{\omega^2T^2m}\left[X-X'-g(T)\left( \frac{x+x'}{2}\right) \right]\right\}\cos\omega t\nonumber\\
&+
\frac{g}{T\sin\omega T}\left\{ x-x'\cos\omega T-(\cos\omega T-1)\frac{gM_{\it eff}}{\omega^2T^2m}\left[X-X'-g(T)\left( \frac{x+x'}{2}\right) \right]\right\}\sin\omega t\nonumber\\
&+
\frac{1}{T}\left(1-\frac{g^2M}{\omega^2T^2m}\right)\frac{M_{\it eff}}{M}\left[X-X'-g(T)\left(\frac{x+x'}{2} \right)\right].
\end{align}
\end{widetext}
With Eq.~\eqref{cl-particle}, the classical time average of the position of the particle is easily obtained:
\begin{widetext}
\begin{align}\label{xbar_clSHO}
\bar x_{cl}=\left(\frac{1-\cos\omega T}{\omega T\sin\omega T} \right)(x+x')+\frac{gM_{\it eff}}{m\omega^2T^2}\left[X-X'-g(T)\left( \frac{x+x'}{2}\right) \right]\left(\frac{g(T)}{g}-1 \right).
\end{align}
\end{widetext}
Then in the expression~\eqref{exact-class-opSHO} for the matrix elements of the class operators we can identify~$Z=\bar x_{cl}$.   For the matrix elements of the class operators the limit~$\hbar\rightarrow 0$ yields
\begin{align}
\!\!\!\!\left\langle {x,X}\right|\widehat C_{\alpha}\left| {x',X'} \right\rangle \approx \mathlarger{e}_{\mathsmaller{\Delta}_{\alpha}}\left(\bar x_{cl}\right)
\left\langle {x, X} \right|{\rm e}^{-\frac{\rm i}{\hbar}\hat HT}\left| {x'.X'} \right\rangle.
\end{align}

This result is similar to the one obtained earlier for the free particle, where~$\bar x_{cl}$ is now given by the result in Eq.~\eqref{xbar_clSHO} and~$H$ is given by~\eqref{H_SHO}.  In the classical limit, decoherence is obtained for the case of a harmonic oscillator coupled to a von Neumann apparatus.  

The explicit form for the matrix elements of the decoherence functional in this case can easily be obtained from the expression~\eqref{decoher-funct} used for the free particle and by
 the exact expression~\eqref{exact-class-opSHO} for the matrix elements of the class operators with the lengths~$Z$ and~$\ell$ given by the expressions~\eqref{Z-SHO} and~\eqref{l-SHO} respectively.  The integral expression for the exact matrix elements of the decoherence functional is
 \begin{widetext}
 \begin{align}\label{DECOHERENCE-functionalSHO}
& \left\langle{x,X}\right|\widehat C_{\alpha}^{\dagger}\widehat C_{\alpha'}\left| {x',X'} \right\rangle=\frac{m\omega}{2\pi\hbar\sin\omega T}\frac{M_{\it eff}}{2\pi\hbar T}{\rm exp}\left[ \frac{{\rm i}m\omega}{2\hbar \sin\omega T}(x'^2-x^2)\cos\omega T\right] \nonumber\\
 &\qquad\qquad\qquad\times
 {\rm exp}\left\{\frac{{\rm i}M_{\it eff}}{2\hbar T}\left[ X'^2-X^2+g(T)\left( X'x'-Xx\right)
 +
 \frac{g(T)^2}{4}(x'^2-x^2)\cos\omega T\right] \right\}\nonumber\\
 &\times
 \int_{-\infty}^\infty\int_{-\infty}^\infty{\rm d}y\,{\rm d}Y{\rm exp}\left[\frac{{\rm i}m\omega}{\hbar\sin\omega T}y(x-x')\right]{\rm exp}\left\{\frac{{\rm i}M_{\it eff}}{2\hbar T}\!\left[2Y(X-X')+g(T)(x-x')Y-g(T)y(X-X')\right] 
 \right\}\nonumber\\
 &\qquad\qquad\qquad\times
 {\mathlarger E}^*_{\Delta_{\alpha}}\left(\bar x_{cl}(x,y,Y,X)\right)
{\mathlarger E}_{\Delta_{\alpha'}}\left(\bar x_{cl}(x',y,Y,X')\right),
\end{align}
\end{widetext}
where the average~$\bar x_{cl}$ of the position of the particle over the classical path is given by Eq,~\eqref{xbar_clSHO}, and the coupling constant~$g(T)$ is defined in Eq.~\eqref{g(T)}.  When we take the classical limit we obtain the same expression as in Eq.~\eqref{class-lim} with~$S_{cl}$ being the action evaluated at the classical path for the harmonic oscillator-apparatus.
 
 As was found in the case of the free particle, there is no decoherence when we are away from the classical region.
 

\section{General Potential. Decoherence in the classical limit}

    In this section we will consider some general arguments for decoherence in the classical limit when a general potential~$V(x,t)$ acts on the particle. The matrix elements of the class operators are given by the expression~\eqref{class-operator}, where the action is defined in Eq.~\eqref{S1} with the proviso that the coupling function in the interaction~\eqref{Hi} is a constant.
    
      The matrix elements of the class operators are
\begin{align}
\left\langle {x,X} \right|\widehat C_{\alpha}\left| {x',X'} \right\rangle&=
\int_U\delta X(t)
\int_u\delta x(t)\nonumber\\
&\times
\mathlarger{e}_{\mathsmaller{\Delta}_{\alpha}}(\bar x[x(t)]){\rm e}^{\frac{\rm i}{\hbar} S[X(t),x(t)]}.
\end{align}
The dominant contribution to the path integral as~$\hbar\rightarrow 0$ comes from the classical paths~$x_{cl}(t),\,X_{cl}(t)$ for the particle and the pointer respectively, and these paths make the action~$S[x(t),X(t)]$ an extremum. The functional~$\mathlarger{e}_{\mathsmaller{\Delta}_{\alpha}}(\bar x[x(t)])$ is to be evaluated at the classical path for the system.  Thus,
\begin{align}\label{class-op-approx}
\left\langle {x,X} \right|&\widehat C_{\alpha}\left| {x',X'} \right\rangle=\mathlarger{e}_{\mathsmaller{\Delta}_{\alpha}}\bar x [x_{cl}(t,x,x',X,X')]\nonumber\\
&\times
K(x,X,T;x',X',0)+\epsilon(x,X;x',X'),
\end{align}
where~$K(x,X,T;x',X',0)$ is the propagator of the system defined by the unrestricted sum over paths~\cf Eq.~\eqref{2D-pathint}.  Here the quantity~$ \epsilon(x,X;x',X')$ is a function of the coordinates.  This function goes to zero when~$\hbar\rightarrow 0$, and~$x_{cl}(t,x,x',X,X')$ is the classical path of the particle as the particle moves from~$x'$ to~$x$ during time~$T$.  The pointer is also moving as a classical particle from~$X'$ to~$X$, also during the time~$T$.
The matrix elements of the decoherence functional can be expressed as
\begin{align}\label{exact-dec-func}
\left\langle{x,X}\right|&\widehat C_{\alpha}^{\dagger}\widehat C_{\alpha'}\left| {x',X'} \right\rangle=\iint{\rm d}Y{\rm d}y\left\langle {y,Y}\right|\widehat C_{\alpha}\left| {x,X} \right\rangle^*\nonumber\\
&\!\!\!\!\!\times
\left\langle {y,Y}\right|\widehat C_{\alpha'}\left| {x',X'} \right\rangle=\int_{-\infty}^\infty{\rm d} Y\int_{-\infty}^\infty{\rm d}y\nonumber\\
&\!\!\!\!\!\!\times
\int\int\delta x(t)\delta X(t)\int\int\delta x''(t')\delta X''(t')\nonumber\\
&\!\!\!\!\!\!\times
\mathlarger{e}_{\mathsmaller{\Delta}_{\alpha}}\left(\bar x[x(t)]\right)\mathlarger{e}_{\mathsmaller{\Delta}_{\alpha'}}\left(\bar x[x''(t')]\right)\nonumber\\
&\!\!\!\!\!\!\times
{\rm exp}\left\{\frac{{\rm i}}{\hbar}\left\{S[x''(t'),X''(t')]-S[x(t),X(t)]\right\}\right\}.
\end{align}
The path integrals are over paths~$x(t),\,X(t)$ and~$x''(t'),\,X''(t')$. 

 When~$\hbar\rightarrow 0$ the exponential oscillates very rapidly and the main contribution to the double path integral comes from the classical paths for the particle and for the pointer.  That is,
\begin{widetext}
\begin{align}\label{decoherent-func-cl}
\left\langle{x,X}\right|\widehat C_{\alpha}^{\dagger}\widehat C_{\alpha'}\left| {x',X'} \right\rangle\sim\iint&{\rm d}y\,{\rm d}Y\mathlarger{e}_{\mathsmaller{\Delta}_{\alpha}}\left(\bar x [x_{cl}(t,y,x,Y,X)]\right)\mathlarger{e}_{\mathsmaller{\Delta}_{\alpha}}\left(\bar x [x_{cl}(t,y,x',Y,X']\right)\nonumber\\
&\times
{\rm exp}\left\{\frac{{\rm i}}{\hbar}\left[S_{cl}(y,Y,x',X')-S_{cl}(y,Y,x,X) \right]\right\}.
\end{align}
\end{widetext}

This expression also follows by direct substitution of the approximate expression~\eqref{class-op-approx} for the matrix elements of the class operators into the expression for the matrix elements of the decoherence functional
\begin{align}
\left\langle{x,X}\right|\widehat C_{\alpha}^{\dagger}\widehat C_{\alpha'}\left| {x',X'} \right\rangle\!=\!\!\iint&{\rm d}Y{\rm d}y\left\langle {y,Y}\right|\widehat C_{\alpha}\left| {x,X} \right\rangle^*\nonumber\\
&\!\!\!\!\!\times
\left\langle {y,Y}\right|\widehat C_{\alpha'}\left| {x',X'} \right\rangle.
\end{align}
In Eq.~\eqref{decoherent-func-cl} when~$x\neq x'$ and~$X\neq X'$ the exponent in~\eqref{decoherent-func-cl} vanishes and so does the factor multiplying the exponential because of the exclusive ranges of~$\mathlarger{e}_{\mathsmaller{\Delta}_{\alpha}}$ and~$\mathlarger{e}_{\mathsmaller{\Delta}_{\alpha'}}$. We further assume that for any given value of~$Y$ the exponent in~\eqref{decoherent-func-cl} is not constant in any subinterval of the integration over~$y$. Similarly for the integration over~$Y$ when~$y$ is kept constant. Under these conditions the Riemann-Lebesgue lemma~\cite{Bender} shows that the expression~\eqref{decoherent-func-cl} vanishes when we take the limit~$\hbar\rightarrow 0$.  Thus we obtain decoherence in the classical limit.

\section{Conclusion}

We have investigated the decoherence of coarse-grained histories of a particle coupled to a von Neumann apparatus.   As posited in Ref.~\cite{FR}, realistic measurement situations must take into account the finite time over which they take place, as well as the localization of the particle to a position interval~$\Delta$.  This has naturally lead  to the consideration of spacetime coarse-grainings in the present paper.

The system under consideration is a particle coupled to a von Neumann apparatus that measures the position of the particle for the case that the dimensionless coupling function in the interaction between the particle and the apparatus is constant. We have studied the system particle-apparatus in the context of the Gell-Mann and Hartle quantum mechanics of closed systems. The coarse-grained histories that we have considered are such that
the real line of all possible values of the time average of the position~$\bar x$ of the particle has been divided into equal length intervals:
\begin{equation*}
\Delta_{\alpha}=\big(\,\bar x_{\alpha}-\delta/2\,,\,\bar x_{\alpha}+\delta/2\,\big],~~{\rm with}~\delta>0.
\end{equation*}
A class~$c_{\alpha}$ has been defined by the set of all particle paths such that, for each path in the class, the time average of the position of the particle is within the interval~$\Delta_{\alpha}$.
In this way  the set~$C$ of all paths in the time interval~$[0,T]$  has been divided into an exhaustive set of mutually exclusive classes~$c_{\alpha}$. For each path in a class, the pointer in the apparatus indicates a time average for the position of the particle such that~$\bar x\in\Delta_{\alpha}$.

 The exact class operators for this system have been evaluated as well as the exact decoherence functional for the two cases of a free particle and for a harmonic oscillator. This shows that these coarse-grained histories do not decohere, not even when initially either the particle or the pointer is in an eigenstate of position.   That is, probabilities cannot be assigned for these coarse-grained histories in this case.  Decoherence is obtained only when the classical limit is taken.
 
 As found in Ref.~\cite{FR-2}, decoherence follows in the quantum regime for certain initial states of the system only when the average is taken to be equal to the arithmetic average of the initial position and final position of the particle.  Of course, this represents a coarser graining than the one utilized in the current work.
 
   In conclusion detailed calculations have been carried out for the evaluation of the matrix elements in the position basis for the class operators and the decoherence functional for the two simple cases of a free particle, and a harmonic oscillator.  The indication of the apparatus at the end of the measurement correlates with the time average of the position of the particle taken during the time  the measurement takes place.  The classical limit of the matrix elements for both the class operators and the decoherence functional, has been obtained, and  we have verified that decoherence is obtained in this limit.   Qualitative arguments for decoherence in the classical limit have also ben presented for the case of a general  potential acting on the particle.  In general, decoherence for spacetime histories is to be expected in the classical limit in the quantum mechanics of closed systems.

\bibliography{CDCH_hist}

\begin{thebibliography}{13}%
\makeatletter
\providecommand \@ifxundefined [1]{%
 \@ifx{#1\undefined}
}%
\providecommand \@ifnum [1]{%
 \ifnum #1\expandafter \@firstoftwo
 \else \expandafter \@secondoftwo
 \fi
}%
\providecommand \@ifx [1]{%
 \ifx #1\expandafter \@firstoftwo
 \else \expandafter \@secondoftwo
 \fi
}%
\providecommand \natexlab [1]{#1}%
\providecommand \enquote  [1]{``#1''}%
\providecommand \bibnamefont  [1]{#1}%
\providecommand \bibfnamefont [1]{#1}%
\providecommand \citenamefont [1]{#1}%
\providecommand \href@noop [0]{\@secondoftwo}%
\providecommand \href [0]{\begingroup \@sanitize@url \@href}%
\providecommand \@href[1]{\@@startlink{#1}\@@href}%
\providecommand \@@href[1]{\endgroup#1\@@endlink}%
\providecommand \@sanitize@url [0]{\catcode `\\12\catcode `\$12\catcode
  `\&12\catcode `\#12\catcode `\^12\catcode `\_12\catcode `\%12\relax}%
\providecommand \@@startlink[1]{}%
\providecommand \@@endlink[0]{}%
\providecommand \url  [0]{\begingroup\@sanitize@url \@url }%
\providecommand \@url [1]{\endgroup\@href {#1}{\urlprefix }}%
\providecommand \urlprefix  [0]{URL }%
\providecommand \Eprint [0]{\href }%
\providecommand \doibase [0]{http://dx.doi.org/}%
\providecommand \selectlanguage [0]{\@gobble}%
\providecommand \bibinfo  [0]{\@secondoftwo}%
\providecommand \bibfield  [0]{\@secondoftwo}%
\providecommand \translation [1]{[#1]}%
\providecommand \BibitemOpen [0]{}%
\providecommand \bibitemStop [0]{}%
\providecommand \bibitemNoStop [0]{.\EOS\space}%
\providecommand \EOS [0]{\spacefactor3000\relax}%
\providecommand \BibitemShut  [1]{\csname bibitem#1\endcsname}%
\let\auto@bib@innerbib\@empty
\bibitem [{\citenamefont {Hartle}(1991{\natexlab{a}})}]{H1}%
  \BibitemOpen
  \bibfield  {author} {\bibinfo {author} {\bibfnamefont {J.~B.}\ \bibnamefont
  {Hartle}},\ }\href@noop {} {\bibfield  {journal} {\bibinfo  {journal} {Phys.
  Rev. D}\ }\textbf {\bibinfo {volume} {44}},\ \bibinfo {pages} {3173}
  (\bibinfo {year} {1991}{\natexlab{a}})}\BibitemShut {NoStop}%
\bibitem [{\citenamefont {Hartle}(1991{\natexlab{b}})}]{H2}%
  \BibitemOpen
  \bibfield  {author} {\bibinfo {author} {\bibfnamefont {J.~B.}\ \bibnamefont
  {Hartle}},\ }in\ \href@noop {} {\emph {\bibinfo {booktitle} {Quantum
  Cosmology and Baby Universes}}},\ Vol.\ \bibinfo {volume} {Proceedings of the
  1989 Jerusalem Winter School for Theoretical Physics},\ \bibinfo {editor}
  {edited by\ \bibinfo {editor} {\bibfnamefont {S.}~\bibnamefont {Coleman}},
  \bibinfo {editor} {\bibfnamefont {J.~B.}\ \bibnamefont {Hartle}}, \bibinfo
  {editor} {\bibfnamefont {T.}~\bibnamefont {Piran}}, \ and\ \bibinfo {editor}
  {\bibfnamefont {S.}~\bibnamefont {Weinberg}}}\ (\bibinfo  {publisher} {World
  Scientific},\ \bibinfo {address} {Singapore},\ \bibinfo {year} {1991})\ pp.\
  \bibinfo {pages} {65--157}\BibitemShut {NoStop}%
\bibitem [{\citenamefont {Roig}(2014)}]{FR-2}%
  \BibitemOpen
  \bibfield  {author} {\bibinfo {author} {\bibfnamefont {F.~S.}\ \bibnamefont
  {Roig}},\ }\href@noop {} {\bibfield  {journal} {\bibinfo  {journal}
  {arXiv:1405.5961}\ } (\bibinfo {year} {2014})}\BibitemShut {NoStop}%
\bibitem [{\citenamefont {von Neumann}(1955)}]{vonN}%
  \BibitemOpen
  \bibfield  {author} {\bibinfo {author} {\bibfnamefont {J.}~\bibnamefont {von
  Neumann}},\ }\href@noop {} {\emph {\bibinfo {title} {Mathematical Foundations
  of Quantum Mechanics}}}\ (\bibinfo  {publisher} {Princeton University
  Press},\ \bibinfo {address} {Princeton, New Jersey},\ \bibinfo {year}
  {1955})\BibitemShut {NoStop}%
\bibitem [{\citenamefont {Roig}(2013)}]{FR-1}%
  \BibitemOpen
  \bibfield  {author} {\bibinfo {author} {\bibfnamefont {F.~S.}\ \bibnamefont
  {Roig}},\ }\href@noop {} {\bibfield  {journal} {\bibinfo  {journal}
  {arXiv:1209.2445}\ } (\bibinfo {year} {2013})}\BibitemShut {NoStop}%
\bibitem [{\citenamefont {Dowker}\ and\ \citenamefont
  {Kent}(1996)}]{Dowker-Kent}%
  \BibitemOpen
  \bibfield  {author} {\bibinfo {author} {\bibfnamefont {F.}~\bibnamefont
  {Dowker}}\ and\ \bibinfo {author} {\bibfnamefont {A.}~\bibnamefont {Kent}},\
  }\href@noop {} {\bibfield  {journal} {\bibinfo  {journal} {Journal of
  Statistical Physics}\ }\textbf {\bibinfo {volume} {82}},\ \bibinfo {pages}
  {1575} (\bibinfo {year} {1996})}\BibitemShut {NoStop}%
\bibitem [{\citenamefont {Schlosshauser}(2005)}]{Schlosshauser}%
  \BibitemOpen
  \bibfield  {author} {\bibinfo {author} {\bibfnamefont {M.}~\bibnamefont
  {Schlosshauser}},\ }\href@noop {} {\bibfield  {journal} {\bibinfo  {journal}
  {Reviews of Modern Physics}\ }\textbf {\bibinfo {volume} {76}},\ \bibinfo
  {pages} {1267} (\bibinfo {year} {2005})}\BibitemShut {NoStop}%
\bibitem [{\citenamefont {Dowker}\ and\ \citenamefont
  {Halliwell}(1992)}]{Dowker-Halliwell}%
  \BibitemOpen
  \bibfield  {author} {\bibinfo {author} {\bibfnamefont {H.~F.}\ \bibnamefont
  {Dowker}}\ and\ \bibinfo {author} {\bibfnamefont {J.~J.}\ \bibnamefont
  {Halliwell}},\ }\href@noop {} {\bibfield  {journal} {\bibinfo  {journal}
  {Physical Review D}\ }\textbf {\bibinfo {volume} {46}},\ \bibinfo {pages}
  {1580} (\bibinfo {year} {1992})}\BibitemShut {NoStop}%
\bibitem [{\citenamefont {Halliwell}(1995)}]{Halliwell}%
  \BibitemOpen
  \bibfield  {author} {\bibinfo {author} {\bibfnamefont {J.~J.}\ \bibnamefont
  {Halliwell}},\ }\href@noop {} {\bibfield  {journal} {\bibinfo  {journal}
  {Annals of the New York Academy of Sciences}\ }\textbf {\bibinfo {volume}
  {755}},\ \bibinfo {pages} {726} (\bibinfo {year} {1995})}\BibitemShut
  {NoStop}%
\bibitem [{\citenamefont {Roig}(2006)}]{FR}%
  \BibitemOpen
  \bibfield  {author} {\bibinfo {author} {\bibfnamefont {F.~S.}\ \bibnamefont
  {Roig}},\ }\href@noop {} {\bibfield  {journal} {\bibinfo  {journal} {Phys.
  Rev. A}\ }\textbf {\bibinfo {volume} {73}},\ \bibinfo {pages} {042106}
  (\bibinfo {year} {2006})}\BibitemShut {NoStop}%
\bibitem [{\citenamefont {Bosse}\ and\ \citenamefont {Hartle}(2005)}]{B-H}%
  \BibitemOpen
  \bibfield  {author} {\bibinfo {author} {\bibfnamefont {A.~W.}\ \bibnamefont
  {Bosse}}\ and\ \bibinfo {author} {\bibfnamefont {J.~B.}\ \bibnamefont
  {Hartle}},\ }\href@noop {} {\bibfield  {journal} {\bibinfo  {journal}
  {Physical Review A}\ }\textbf {\bibinfo {volume} {72}},\ \bibinfo {pages}
  {022105} (\bibinfo {year} {2005})}\BibitemShut {NoStop}%
\bibitem [{\citenamefont {Feynman}\ and\ \citenamefont
  {Hibbs}(1965)}]{Feynman}%
  \BibitemOpen
  \bibfield  {author} {\bibinfo {author} {\bibfnamefont {R.~P.}\ \bibnamefont
  {Feynman}}\ and\ \bibinfo {author} {\bibfnamefont {A.~R.}\ \bibnamefont
  {Hibbs}},\ }\href@noop {} {\emph {\bibinfo {title} {Quantum Mechanics and
  Path Integrals}}}\ (\bibinfo  {publisher} {McGraw-Hill},\ \bibinfo {address}
  {New York},\ \bibinfo {year} {1965})\BibitemShut {NoStop}%
\bibitem [{\citenamefont {Bender}\ and\ \citenamefont {Orszag}(1978)}]{Bender}%
  \BibitemOpen
  \bibfield  {author} {\bibinfo {author} {\bibfnamefont {C.~M.}\ \bibnamefont
  {Bender}}\ and\ \bibinfo {author} {\bibfnamefont {S.~A.}\ \bibnamefont
  {Orszag}},\ }\href@noop {} {\emph {\bibinfo {title} {Advanced Mathematical
  Methods for Scientists and Engineers}}}\ (\bibinfo  {publisher}
  {McGraw-Hill},\ \bibinfo {year} {1978})\BibitemShut {NoStop}%
\end{thebibliography}%

\end{document}